\documentclass[useAMS,usenatbib]{mn2e}
\usepackage[varg]{txfonts}
\usepackage[utf8]{inputenc}
\usepackage{colortbl}
\def\lsim{\lower.5ex\hbox{$\; \buildrel < \over \sim \;$}}
\def\gsim{\lower.5ex\hbox{$\; \buildrel > \over \sim \;$}}
\def\be{\begin{equation}}
\def\bea{\begin{eqnarray}}
\def\eea{\end{eqnarray}}
\def\ee{\end{equation}}

\def\mdtj{{\dot {\cal M}}}

\def\bc{\begin{center}}
\def\ec{\end{center}}

\def\etal{{\em et al.}}
\def\ie{{\em i.e.}}
\def\ep{{e^--p^+}}
\def\el{{e^--e^+}}

\def\sk{{\rm sk}}

\def\mbh{M_{\rm B}}

\def\rs{r_{\rm s}}

\def\xsh{x_{\rm sh}}

\def\rs{r_{\rm \small S}}
\def\ep{{{\rm e}^--{\rm p}^+}}

\def\gamt{\gamma_{\rm \small T}}

\def\od{\rm \small D}
\def\msol{M_\odot}

\include{netbib.sty}

\usepackage{graphicx}
\title[Compton driven relativistic jets]
{Radiation driving and heating of general relativistic jets under Compton scattering regime}
\author[Vyas \& Chattopadhyay]
{Mukesh K. Vyas$^{1}$, Indranil Chattopadhyay$^{1}$ \thanks{Email:mukesh.vyas@aries.res.in (MKV);
		indra@aries.res.in (IC)}\\
$^{1}$Aryabhatta Research Institute of Observational Sciences 
(ARIES), Manora Peak, Nainital-263002, India\\
}
\begin{document}
\date{}
\maketitle
\label{firstpage}
\begin{abstract}
Interaction of intense radiation from the underlying accretion disc with steady, general-relativistic jet is studied. The radiation field imparts momentum as well as energy on to the outflowing jet under Compton scattering. As a result, the jet gains momentum and is simultaneously heated up. 
{Jets }can be classified as types A, B and C
according to their base properties. We found { that }A type jets can undergo shock transition. 
It is also shown that,{ in the Compton scattering regime, radiation can drive jets starting with very small thermal energy at the base  (B and C type jets)}. 
{ Such that, radiation can even} accelerate bound matter (generalized Bernoulli
parameter $E<1$) {in the form of} relativistic transonic jets.
This is in stark contrast to radiatively driven jets in { the}
Thomson scattering regime, where transonic jets were obtained only for $E>1$.
We also show that for a given disc luminosity, { jets in the Compton scattering regime}
exhibit a minimum terminal speed, unlike in the Thomson scattering domain.
Further, the impact of accretion disc luminosity and jet plasma composition is studied. The $\ep$ jets are accelerated up to Lorentz factors of about a few, while for lepton dominated jets the minimum Lorentz factor exceeds $10$ for moderate disc luminosities and can go up to few tens for highly luminous discs.
\end{abstract}


\twocolumn
\section[Introduction]{Introduction}
\label{sec1}
\begin {figure}
\begin{center}
 \includegraphics[width=9.cm]{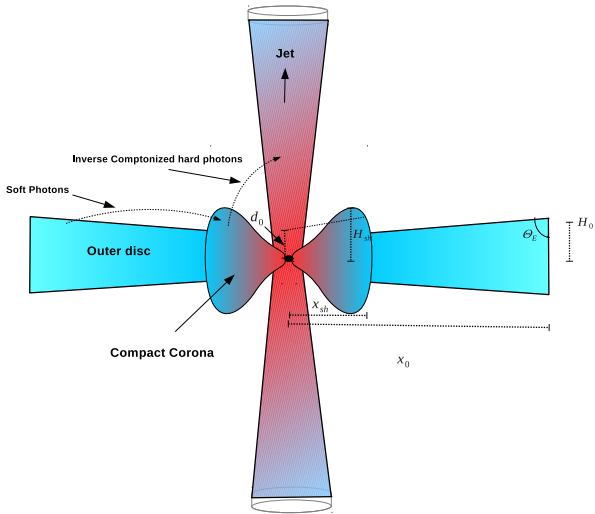}
\vskip -0.1cm
 \caption{Cartoon diagram of disc-jet system. The outer extent and height of the corona
$\xsh$ and $H_{\rm sh}$, the intercept of outer disc on the axis ($d_0$), the outer edge of the
disc $x_0$ are shown. The
funnel of the corona is also shown.}
\label{lab:fig1}
 \end{center}
\end{figure}
Astrophysical jet was first discovered by \cite{c18} in optical wavelength while studying M87.
After the advent of radio astronomy, these jets were studied in { detail} in the later half of the 19th century.
Since then the jets have been recognized 
as ubiquitous astrophysical phenomena associated with various classes of objects like active galactic
nuclei (AGN e.g., M87, 3C 279), young stellar objects (YSO e.g., HH 30, HH 34), X-ray binaries (e. g., SS433, Cyg
X-3, GRS 1915+105, GRO 1655-40).

In black hole (BH) sources, jets can only emerge from accreting matter because BHs are not capable of emitting
matter or radiation. The strong correlation observed between spectral states of accretion disc and jet evolution \citep{gfp03,fgr10,rsfp10}, suggests that { the jets are launched by the disc}.
Further, jets were observed to be launched from a region within few tens of Schwarzschild radii ($r_{\rm s}$)
from the central BH \citep{jbl99,detal12}.
Hence jets are generated from the hot and more active inner region of the accretion disc. These facts make it important to study the impact of thermal pressure as well as the radiation field on the dynamics of the jet.

Along with the development of various accretion discs models \citep{ss73, pw80, nkh97, f87, c89}, several attempts were made to understand the interaction of radiation from these discs with the outflowing jets. 
In this paper, the jets are studied in radiation hydrodynamic (RHD) regime. { The} RHD equations of motion (EoM) were developed by various authors in special relativistic or SR regime \citep{hs76,cn82,mm84,kfm98} and in general relativistic (GR) regime \citep{p06, t07}.

Significant development of the field started from 1980s. 
\citet{sw81} studied particle jets driven by radiation in SR regime.
\citet{o81} showed that the Thomson scattering radiation force increases for hot plasma, which might
result in enhanced radiative driving called `Compton rocket'. However, \citet{p82} played-down the significance of Compton rocket in presence of Compton cooling.
\cite{ftrt85} studied radiatively driven fluid jets in SR regime. They considered Newtonian gravitational potential along with isothermal equation of state (EoS) with which they produced mildly relativistic jets and obtained internal shocks as manifestation of jet geometry. \cite{i89} studied { the }role of radiation drag on particle jets plying through the radiation field { produced by} an infinite Keplerian disc. { He} discovered the upper limit of matter speed to be $0.45c$, where $c$ is the speed of light in vacuum.  \cite{f96} studied particle jets under radiation field considering pseudo-Newtonian potential (pNp) to take care of strong gravity. \citet{fth01} considered a hybrid disc and
produced jets with $\gamt \sim 2$.

Through various numerical \citep{mrc96,dcnm14,lckhr16}
and theoretical studies \citep{cd07,kc13,kcm14,kc17,kscc13,ck16}, it was found that extra thermal gradient force in the { corona} close to the BH automatically generates bipolar outflows.
Driving of jets by radiation from advective disc was also investigated
\citep{cc00a,cc00b,cc02a,cc02b}. In non-relativistic regime, the formalism followed by \citet{cc00a,cc02a} is only correct up to the first order of the flow velocity. 
Full relativistic { transformations} of radiative moments were later considered { and implemented}
for particle jets in SR regime \citep{cdc04,c05}. The disc chosen was of the hybrid type
with two sources of radiation, (i) the Keplerian disc and (ii) the post-shock region of the sub-Keplerian component of the disc.
These authors also showed that relativistic jets ($\gamt \gsim 2$) with impressive collimation
could be achieved for such jets.

Most of the above cited studies were either conducted 
{ in the particle} regime, { that is,} the gas pressure was neglected compared to the radiation pressure,
or, by considering fluids in the non-relativistic regime. 
Vyas et. al. (2015, hereafter \citet{vkmc15}) solved RHD equations of motion for jets
described by relativistic equation of state (EoS).
The authors considered special relativistic space-time and the gravity was mimicked by a pNp which was initially proposed by
\citet{pw80}. It is well known that, for hot flows radiation driving becomes in-effective, and yet
\citet{vkmc15} showed that jets can be { accelerated to} relativistic terminal speeds.
The strong temperature gradient drives the jet just above the base, and thereafter
radiation driving takes over. In fact, the thermal driving is so strong that it accelerates { the} jet to a speed of { about 10\% of the speed of light within first few Schwarzschild
radii above the disc. In this region} the radiation field actually decelerates the jet due to an effect called radiation drag, { beyond which the jet is accelerated }by the radiation field.

There were two limitations of \citet{vkmc15} --- (i) combining SR with a gravitational potential and, (ii)
consideration of conical flow geometry. We addressed these two issues separately in general { relativistic analysis}.
We showed the formation of moderately strong internal shocks as a direct fall out of non-conical geometry
(Vyas \& Chattopadhyay 2017, hereafter \citet{vc17}). And then radiation driving of jets in curved
geometry was considered for flow described by fixed adiabatic index ($\Gamma$) EoS \citep{vc18a}
as well as with relativistic EoS (Vyas \& Chattopadhyay 2018b, hereafter \citet{vc18b}).
We showed that, considering jets in SR regime with an ad hoc gravitational potential produces un-physically hot flow, producing additional thermal gradient push. In other words, because
jets are supposed to be launched close to the compact object, consideration of GR is important.

To simplify the analysis in most of the works cited above, the interaction of disc radiation with jet matter was considered in the elastic scattering regime. Under { this} assumption, { the} radiation field { only} transfers momentum to the jet but { there is no energy transfer between them}. { If} the radiation is of the order of $\lsim \mbox{ few }\times 10$ KeV, then elastic scattering
cross section (i. e., Thomson scattering cross section) is applicable, but if higher energy photons impinge
on electrons then Thomson scattering cross section is untenable.
There were very few studies on Compton driving of outflows. \citet{qp85,tnc86} considered radiatively driven winds
accompanied by energy exchange between radiation and matter mostly in the optically thick regime.
There were initial apprehension about `a severe' limit on driving a jet in the optically thin regime
because of the presence of radiation drag \citep{i89}, however later it was shown by a number of authors,
that if the bulk of intense radiation comes from the inner part of the accretion disc, radiation drag
ceases to be a major obstacle for jet driving \citep{cdc04,c05} at a distance of about $100$ Schwarzschild radius above the disc plane. And at such distances the radiation field remains significant enough to drive the jet matter.

{ In the current paper}, we consider radiatively driven jets in curved space-time using a general scattering cross section \citep{by76,p83}, which enables energy as well as momentum transfer from radiation to the jet matter. 
The thermodynamics of the jet material is described by a relativistic EoS \citep{c08,cr09}, and the radiation
moments were computed by considering the effects of space-time curvature \citep{b02}.
In the elastic scattering regime \citep{vc18b}, the radiatively driven jet solutions were 
primarily of two types (i) { the jets, that are launched with hot base and higher speeds} and (ii) { the jets are launched with moderately hot base but with almost negligible base speeds}. The second {type of solutions were} due to the geometrically thick corona.
In this paper the height of the inner corona is adopted from \citet{kfm98}, and the corona height cannot
increase beyond a limit. Therefore, it is indeed intriguing to find out, how would the jet solutions
change in light of the two modifications considered, namely, Compton scattering cross section and
modified corona size.
\citet{vc18b} also obtained radiatively driven internal shocks. Do such solutions survive in the Compton regime? How would the jet solutions be modified
if the composition of the flow is varied, these { are some of the questions that are discussed} in this paper.   

In next section we present detailed mathematical formalism including EoM and estimation of radiation field. Then briefly discuss about methods of obtaining results in section \ref{sec:method}. We describe results of the study in section \ref{sec:res} and conclude the paper in section \ref{sec:con}. Description of used relativistic EoS is given in Appendix (\ref{app:EoS}) while detailed method of calculating radiation field is given in Appendix (\ref{app:rad_mom}).

\section{Mathematical formalism}
\label{sec:mat}
\subsection{Space-time metric, unit system and assumptions made in the study}
We consider a non-rotating black hole described by a Schwarzschild metric:
\bea
ds^2=-g_{tt} c^2dt^2+
g_{rr}dr^2+g_{\theta \theta}d{\theta}^2+g_{\phi \phi}d\phi^2 
=-\left(1-\frac{2G\mbh}{c^2r} \right)c^2dt^2 \nonumber \\ 
+\left(1-\frac{2G\mbh}{c^2r}\right)^{-1}dr^2 
+r^2d{\theta}^2+r^2\sin^2{\theta}d\phi^2
\label{metric.eq}
\eea
Here, $r$, $\theta$, $\phi$ and $t$ are space-time coordinates.
$\mbh$ and $G$ are the black hole (BH) mass and the
universal constant of gravitation, respectively. In this paper, we have used geometric units $G=\mbh=c=1$, such
that the event horizon or Schwarzschild radius is $\rs=2$. The jet is assumed to be around the axis of symmetry of the underlying accretion disc. The jet is stationary and
axis-symmetric (i. e., $\partial/\partial t=\partial/\partial \phi=0$). Since jets are collimated, we consider the jet to be conical (i. e., the cross-section ${\cal A}\propto r^2$) with a narrow opening angle. 
In this paper we do not consider exact launching mechanism of jets from the accretion disc, instead the accretion disc acts only as the source of radiation. The accretion disc is around the equatorial plane.    
The inner part of accretion disc has a geometrically thick corona and the expression of the height of the corona ($H_{\rm sh}$) is given as \citep{kfm98} 
\be
H_{\rm sh}=H^*\left(1-\sqrt{\frac{2}{\xsh}}\right)
\ee
Here $H^*$ and $\xsh$ are upper limit of the corona height and horizontal extent of the corona respectively. The expression
of corona cross-section is different from \citet{vc18b}, where the corona was assumed to be thick. In current 
paper, if the horizontal extent of the corona is large, then the corona is geometrically slim
($H_{\rm sh}/x_{\rm sh}< 1$). 
A typical cartoon diagram of the assumed system is given in Fig (\ref{lab:fig1}) which shows a bipolar radial jet coming out of nearby region inside the coronal funnel. Outer portion of the disc is also shown.  The cartoon
of the disc-jet system presupposes the jet is launched from the inner part of the disc, although not explicitly
computed from first principle.

\subsection{Radiation hydrodynamic equations governing the dynamics of relativistic fluids }
The energy-momentum tensor for the matter ($T^{\mu \nu}_M$) and radiation ($T^{\mu \nu}_R$)
is given by
\begin{eqnarray}
T^{\mu \nu}=T^{\mu \nu}_M+T^{\mu \nu}_R;\mbox{ where }T^{\mu \nu}_M=(e+p)u^{\mu}u^{\nu}+pg^{\mu \nu};
\nonumber \\
\mbox{ and } T^{\mu \nu}_R={\int}I ~l^{\mu}l^{\nu}d{\Omega},
\label{tmunu.eq}
\end{eqnarray}
Here,  $u^{\mu}$ are the components of four velocity,
$l^{\mu}$s are directional derivatives, $I$ is the frequency integrated specific intensity
of the radiation field 
and $d\Omega$ is the differential solid angle subtended by a source point on the accretion disc surface to the field point on the jet axis. The assumption of conical outflow along the axis of symmetry of the accretion disc,
implies that the only significant component of four velocity is $u^r$.

The equations of motion are given by
\begin{equation}
T^{\mu \nu}_{;\nu}=0~~~~  \mbox{and} ~~~~
(\rho u^{\nu})_{;\nu}=0,
\label{eqnmot.eq}
\end{equation}



The momentum balance equation obtained under the present set of assumptions is along the radial direction,
\begin{equation}
u^r\frac{du^r}{dr}+\frac{1}{r^2}=-\left(1-\frac{2}{r}+u^ru^r\right)
\frac{1}{e+p}\frac{dp}{dr}+{\rho}_e\frac{{\sqrt{g^{rr}}\gamma^3}
}{(e+p)}{\Im}^r
\label{eu1con.eq}
\end{equation}

Here, $\gamma=-u_tu^t=(1-v^2)^{-1/2}$ is the bulk Lorentz factor of the jet, where $v=\sqrt{(-u_ru^r/u_tu^t)}$
is the three velocity in the radial direction. It means $u^r=\sqrt{g^{rr}}{\gamma}v$.
The total lepton mass density is given by $\rho_e$ and
${\Im}^r$ is the momentum imparted onto the jet plasma by the radiation field of the accretion disc
and is given by
\be
{\Im}^r=\frac{\sigma}{m_e}\left[(1+v^2){\cal R}_1-v
\left(g^{rr} {\cal R}_0+\frac{{\cal R}_2}{g^{rr}}\right)\right]
\label{radcontrib.eq}
\ee
The first three moments of the specific intensity of the radiation field
are ${\cal R}_0,~{\cal R}_1$ and
${\cal R}_2$ and are identified
as the radiation energy density, the flux and the pressure, respectively.
However, similar to \citet{vc18b}
we would refer $R_0=\sigma_T{\cal R}_0/(m_e),~R_1=\sigma_T{\cal R}_1/(m_e)$ and $R_2=\sigma_T{\cal R}_2/(m_e)$ 
as respective radiative moments. The detailed method of estimation of radiative moments is given in appendix \ref{app:rad_mom} \citep[see also][]{vc18b}. The negative terms arise for optically thin medium, and as long as the jet `sees' the accretion disc as an extended radiator. These negative terms are called `radiation drag' terms and arise because of the anisotropic nature of the radiation field and are stronger near the disc surface.
The scattering cross section $\sigma$ is
given as \citep{by76,p83}, 
\be
\sigma= \chi_c\sigma_T=\left[\frac{1}{1+\left(\frac{T_e}{4.5 \times 10^8}\right)^{0.86}}\right]\sigma_T,
\ee
where, $\sigma_T$ is Thomson scattering cross section. $\chi_c$ accounts for Compton process (i.e., energy exchange between radiation and matter) and is $<$ 1. $T_e$ is electron temperature in physical units.
It is approximated as a function of $N$
\citep{kc14,sc18}
$$
\Theta_e=\frac{kT_e}{m_ec^2}=-\frac{2}{3}+\frac{1}{3}\sqrt{\left[4-2\left(\frac{2N-3}{N-3}\right)\right]}
$$
Although the form of equation \ref{eu1con.eq} is similar to the one in \citet{vc18b}, however the 
difference is in the expression of $\sigma$. 

The first law of thermodynamics, or energy equation ($u_{\alpha}T^{\alpha \beta}_{M_{;\beta}}=-u_{\mu}T^{\mu \nu}_{R_{;\nu}}$) is obtained as,
\begin{equation}
\frac{de}{dr}-\frac{e+p} {\rho}\frac{d\rho}{dr}=-\frac{\gamma \rho_e(1-\chi_c)R_t}{\sqrt{g^{rr}}},
\label{en1con.eq}
\end{equation} 
Here $R_t$ is radiative contribution representing energy exchange between imparted radiation and fluid
\citep[same as the heating term of][]{p06}: 
\be
R_t=\left[\frac{g^{rr} {R}_0}{v}+\frac{v {R}_2}{g^{rr}}-2{R}_1\right]
\ee 
Integrating the conservation of mass flux (the second of equation \ref{eqnmot.eq}),
we obtain the mass outflow rate 
\begin {equation}
{\dot M}_{\rm o}=\rho u^r \cal A
\label{mdotout.eq}
\end {equation}
Here ${\cal A}$ is the cross-section of the jet. Since the jet is transonic and collimated, we assume that the cross section to be conical (${\cal A} \propto r^2$). However, the radiation supplies energy to the jet and makes it hotter, which might raise the apprehension that assumption of conical jet may not hold. We discuss this in greater details in section \ref{sec:con} and Appendix \ref{cross_section} and we show that 
the assumption is reasonable.
In \citet{vc18b}, elastic scattering assumption rendered $R_t=0$ and therefore integrating
equation (\ref{en1con.eq}) with the help of the EoS (equation \ref{eos.eq}), we obtained the adiabatic relation between $\Theta$ and $\rho$
\citep{kscc13}. Replacing $\rho$ of the adiabatic relation, into equation (\ref{mdotout.eq}) we also obtained the
expression of entropy-outflow rate \citep{vkmc15, vc18b},
\begin{equation}
\mdtj=\mbox{exp}(k_3) \Theta^{3/2}(3\Theta+2)
^{k_1}
(3\Theta+2/\eta)^{k_2}u^r{\cal A},
\label{entacc.eq}
\end{equation}
where, $k_1=3(2-\xi)/4$, $k_2=3\xi/4$, and $k_3=(f-\tau)/(2\Theta)$.
This is also a measure of entropy of the jet and unlike \citet{vkmc15,vc18b}, $\mdtj$ is not a constant
in this paper.

Integrating the first of equation (\ref{eqnmot.eq}) is equivalent to integrating equations (\ref{eu1con.eq}) and (\ref{en1con.eq}) simultaneously and we obtain the generalized, relativistic Bernoulli constant as the
constant of motion,
\bea 
 E=-h u_te^{-X_f},~~~~\mbox{where }
 \nonumber  \\
 X_f=\int dr \frac{\gamma (2-\xi)}{(f+2 \Theta )\sqrt{g^{rr}}}\left[\Im^r-(1-\chi_c)R_t\right]
\label{energy.eq}
\eea



In absence of radiation it is merely $E=E_t=-hu_t$. With the help of equation (\ref{eos2.eq}), equations (\ref{eu1con.eq}) and (\ref{en1con.eq}) can be expressed as gradients of $v$ and $\Theta$
and are given by
\begin{eqnarray}
\label{dvdr.eq}
\gamma^2vg^{rr}r^2\left(1-\frac{a^2}{v^2}\right)\frac{dv}{dr}=a^2\left(\frac{g^{rr}r^2}{\cal A}\frac{d\cal A}{dr}+1\right)-1  \nonumber \\
+\frac{(2-\xi)\gamma r^2 \sqrt{g^{rr}}}{f+2 \Theta}\left[\Im^r-\frac{(1-\chi_c)R_t}{N}\right]
\end{eqnarray}
and

\begin{equation}
\frac{d{\Theta}}{dr}=-\frac{{\Theta}}{N}\left[ \frac{{\gamma}
^2}{v}\left(\frac{dv}{dr}\right)+\frac{1}{\cal A}\frac{d\cal A}{dr}+\frac{1}{g^{rr}r^2}-\frac{(2-\xi)(1-\chi_c)\gamma R_t}{2 \Theta \sqrt{g^{rr}}}
\right]
\label{dthdr.eq}
\end{equation}

\section{Method of obtaining solutions}
\label{sec:method}
\label{Sonic_point_conditions}
Jet solutions are obtained by integrating equations \ref{dvdr.eq} and \ref{dthdr.eq}. 
The jet base is close to the horizon. As it is hot and slow near the base, it is subsonic. 
In this paper the jet base is generally considered to be at $r=r_b=3$ until specified otherwise.
At large distances from the BH, the jet moves with very high speed and is cold and hence it is supersonic.
In other words, the jets are transonic, subsonic branch of the solution passes on to the supersonic one through the sonic point ($r=r_c$), i. e., at $r_c$, $v_c=a_c$. Here suffix $c$ denotes quantities at the
sonic point.
Further, at $r_c$, $dv/dr\rightarrow 0/0$, which enables us to write down the other sonic point condition as
\be
\left[ a^2\left(\frac{g^{rr}r^2}{\cal A}\frac{d\cal A}{dr}+1\right)-1+\frac{(2-\xi)\gamma r^2 \sqrt{g^{rr}}}{f+2 \Theta}\left\{\Im^r-\frac{(1-\chi_c)R_t}{N}\right\}\right]_{r=r_c}=0
\label{sonic2.eq}
\ee

The $dv/dr|_c$ is calculated by employing the L'Hospital's rule at $r_c$ and solving the resulting
quadratic equation of $dv/dr|_c$. The resulting quadratic equation can admit complex root leading to
the spiral type
sonic points, or two real roots. The solutions with two real roots but with opposite signs
are called $X$ or `saddle' type sonic points, while real roots with same sign produces the nodal type sonic point. The
jet solutions passing through X type sonic points are physical. 
So for a given set of flow variables at the jet base, a unique solution will pass through
the sonic point determined by the entropy ${\dot {\cal M}}$ of the flow. For given
values of inner boundary condition, that is, at the jet base  $r_b$, $v_b$ and $\Theta_b$
we integrate equation (\ref{dvdr.eq}) and (\ref{dthdr.eq}), while checking
for the sonic point conditions (equations \ref{sonic2.eq}). A set of $r_b,~v_b, \mbox{ and }\Theta_b$
corresponds to a particular value of the constant of motion $E$. Various combination of $v_b$ and
$\Theta_b$ can give rise to the same $E$, but only a particular value of $\mdtj$ corresponding to the same $E$,
will admit a sonic point. And following the second law of thermodynamics, $\mdtj$ of the transonic solution
is maximum for all global solutions.
We iterate till the
sonic point is obtained, and once it is obtained we continue to integrate outwards
starting from the sonic point using Runge Kutta's $4^{th}$ order method. This process gives us values of $v$ and $\Theta$ along $r$. All other variables such as $a$, $\Gamma$, $N$, $h$, $E_t$ and $E$ are obtained from these two variables.
As explained in detail in \citet{vc18b}, we check for shock transition by conserving fluxes at each point, namely mass flux, momentum flux and energy flux.

\section{Analysis and results}
\label{sec:res}
\subsection{Nature of radiation field}
\begin {figure}
\begin{center}
 \includegraphics[width=10.cm, trim=20 0 50 300,clip]{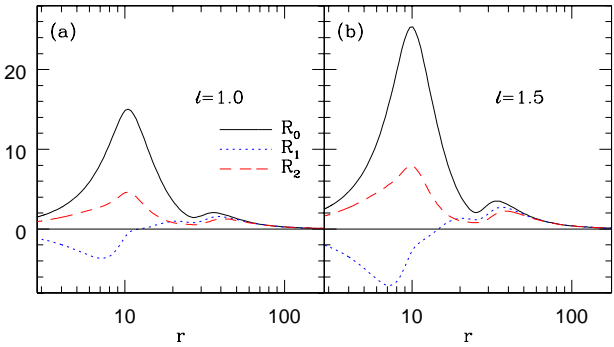}
\vskip -0.5cm
 \caption{Distribution of radiative moments $R_0$ (solid black), $R_1$ (dotted blue) and $R_2$ (red dashed) for (a) $\ell=1.0$ and (b) $\ell=1.5$ along the jet length $r$}
\vskip -0.60cm
\label{lab:rad_mom}
 \end{center}
\end{figure}

We calculate radiative moments as explained in Appendix (\ref{app:rad_mom}). In Fig. (\ref{lab:rad_mom}) we show intensity of radiation field along $r$ by plotting radiative energy density $R_0$ (solid black), $r$ component of radiative flux $R_1$ (dotted blue) and $rr$ component of radiative pressure $R_2$ (red dashed) for various disc luminosities of the accretion disc, $\ell=1.0$ and $1.5$ in panels a and b respectively. These luminosities correspond to accretion rates $\dot{m}=4.62$ and $4.89$ respectively. The radiation field gets weaker as the luminosity decreases. Radiation flux $R_1$ is negative inside the funnel of the corona, which adds to the radiation drag term and resists the jet flow while the flux is positive above the corona, hence it accelerates
the jet. So radiation can accelerate or decelerate and also {heats} up the flow. We will further explain these effects in next section.
\subsection{Flow variables at sonic points}
\begin {figure}
\begin{center}
 \includegraphics[width=8.cm, trim=0cm 0 0cm 0,clip]{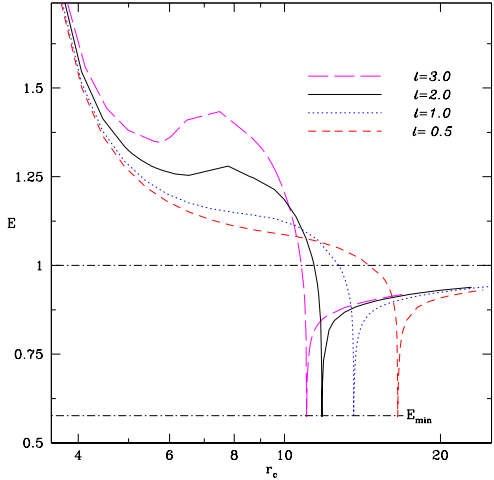}
\vskip -0.5cm
 \caption{Variation of 
 $E$ with $r_c$ for $\ell=3.0$ (long dashed magenta) $\ell=2.0$ (solid black), $\ell=1.0$ (dotted blue) and $\ell=0.5$ (dashed red)}
\label{lab:sonic-prop}
 \end{center}
\end{figure}
As shown before, sonic point analysis is an important aspect of obtaining flow solutions because at sonic point flow speed $v$ equals sound speed $a$, which is essentially a mathematical boundary. Each sonic point corresponds to certain $E$ or equivalently, corresponds to certain jet base parameters like $v_b$ and $\Theta_b$. In Fig. (\ref{lab:sonic-prop}) we plot $E$, for $\ep$ ($\xi=1$) flow for different disc luminosities $\ell=3.0$ (long dashed magenta), $2.0$ (solid black), $1.0$ (dotted blue) and $0.5$ (dashed red). The evolution of $E$ indicates that higher $\ell$ makes the flow more energetic and $E$ becomes non monotonic. 
There are a couple of interesting features that separate the $E-r_c$ curve of this paper with our previous ones \citep{vc17,vc18a,vc18b}, and they are ---
\begin{enumerate}
\item $E$ dips below one.
\item $E$ dips upto a certain minimum value $E_{\rm min}$ and then another
branch is obtained. Although the $r_c$ at $E_{\rm min}$ is different for different luminosity, but
the energy is exactly the same. 
\end{enumerate}
In next section, this phenomena is discussed in details.
 
\subsection{General pattern of solutions and significance of Compton scattering}
\begin {figure}
\begin{center}
 \includegraphics[width=9cm, trim=0 0 0 7cm,clip]{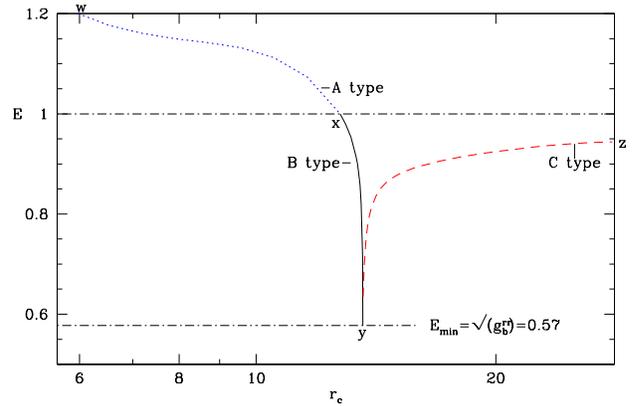}
\vskip -0.5cm
 \caption{Variation of $E$ with $r_c$ for $\ell=1.0$. Depending upon nature of jet base, the solutions are classified in three types: A-type (blue dotted), B-Type (solid black) and C type (red dashed). These are marked in the figure}
\label{lab:gen_sol}
 \end{center}
\end{figure}
 As $E$ is a constant of motion, so Fig. (\ref{lab:sonic-prop}) contains information of all types of jet solutions. Each point on the figure corresponds to certain base variables $v_b$ and $\Theta_b$ corresponding to which
there is a transonic solution. The solutions can be classified into three categories. To show their classification, we again plot $E-r_c$ curve for $\ell=1$ in Fig (\ref{lab:gen_sol}). Three
types of jet solutions are named as A, B, and C. The collective information of base variables lie in the expression of $E$ at the base, {that is},
$E (r_b)=E_b=-h_bu_{tb}=h_b\sqrt{g^{rr}_b}\gamma_b$. { In other words, the radiative contribution enters the Bernoulli expression as the jet propagates along $r$ (equation \ref{energy.eq}).}
Interestingly, energy exchange between the jet and radiation allows even
initially bound matter ($E<1$) to be driven as transonic jets. This is a significantly different result
in comparison to previous studies in the elastic scattering limit where we were restricted with $E>1$ for any transonic solution \citep[see figure 6a of][]{vc18b}. 
We obtained this phenomenon previously in non-relativistic studies
\citep{cc00a,cc00b} but such solutions were not systematically explored and the heating term was an ad hoc one. 

Class A (w-x, dotted blue) and B (x-y, solid black) in Fig. (\ref{lab:gen_sol}) represent sonic point properties for jets starting with the same base ($r_b$). Type C (z-y, red dashed) represents sonic point properties of jets with same $\ell$ but with $r_b>3$. Class A represents high energy jets, and class B were bound matter being driven off as jets starting with the same $r_b$. Both classes B and C were not obtained in the elastic scattering regime i.e., bound matter being driven out as jets.
In addition, class C are jet solutions of bound matter which are suspended at some height above the disc, while for the B class jets, $r_b=3$ is similar to class A. 
At $r_b=3$ the minimum energy matter is obtained when $v_b\rightarrow 0$ and $\Theta_b \ll 1$, i. e. 
\begin{equation}
E_{\rm min}=\sqrt{g^{rr}}; \mbox{ i. e., }\gamma_b \rightarrow 1;~~\&~~ h_b\rightarrow 1 
\end{equation}
At $E_{\rm min}$, B and C class solutions merge. It is precisely for this reason that $E-r_c$
reaches upto $E_{\rm min}$ for any $\ell$ (Figs. \ref{lab:sonic-prop}, \ref{lab:gen_sol}).
In the following we separately discuss each class of solutions.

\subsection{A-type solutions : Hot and fast jet base}
\begin {figure}
\begin{center}
 \includegraphics[width=9cm, trim=0.5cm 0 0 0,clip]{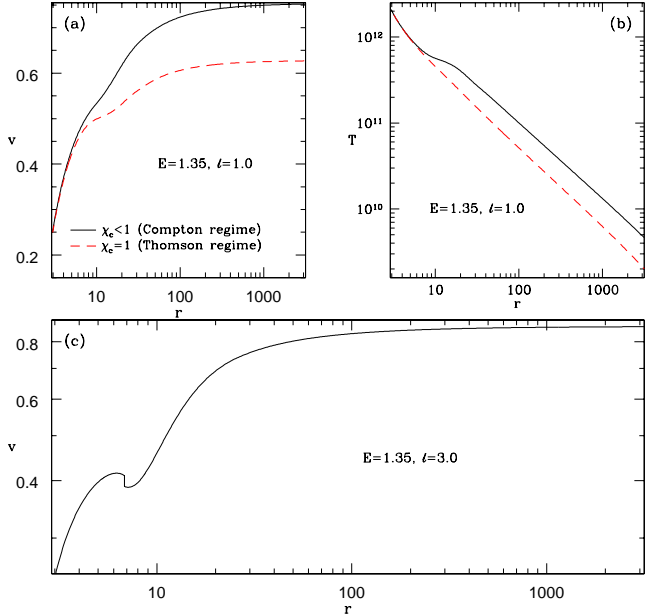}
\vskip -0.5cm
 \caption{Nature of A-type solutions. Variation of (a) jet three velocity $v$ and (b) Temperature $T$ with $r$ for $E=1.35$ and $\ell=1.0$. Solid curve incorporates Compton scattering while dashed red curve considers Thomson scattering. (c)  variation of $v$ for $\ell=3.0$, $E=1.35$.}
\label{lab:A_type}
 \end{center}
\end{figure}
In Figs. (\ref{lab:A_type}a) and (\ref{lab:A_type}b), we plot jet velocity $v$ and temperature $\Theta$ as functions of $r$ for $E=1.35$ and $\ell=1.0$.
The jet in the Compton scattering regime has 
base velocity $v_b=0.14$ and base temperature $T_b\sim2\times10^{12} \rm K$. The sonic point 
is at $r=5.97$, while the terminal speed $v_t=0.69$. Here terminal speeds are 
defined to be $v_t=v|_{r\rightarrow 10^6}$. 
To show the effect of Compton scattering on jet, we 
over-plot the jet solution in Compton scattering regime (solid, black) with that in the Thomson scattering (dashed, red). The jet solution in the Thomson scattering regime is achieved in the present formalism by considering $\chi_c=1$ and these jet solutions are similar to { the ones obtained by \citet{vkmc15} and \citet{vc18b}}. The terminal speed of the jet in the Thomson scattering regime is only $0.52$. 
It is clear that the Compton driven jet is $33\%$ faster compared to a jet under Thomson scattering regime.
The temperature profile of the jet in the elastic scattering regime monotonically decreases, while the temperature
of the Compton jet is not monotonic, because the jet is heated up at around $r \sim 20$ 
where radiative moments peak (Fig \ref{lab:rad_mom}b). It again cools down monotonically with $r$, as radiation field  gets weaker further away. In Fig. (\ref{lab:A_type}c), we plot $v$ as a function of $r$ of a jet with same energy i. e., $E=1.35$, however, it is acted on by radiation characterized by $\ell=3$. For these parameters, the jet accelerates and becomes transonic at the inner sonic point
at $r_c=5.3$ and goes through shock transition under the impact of negative flux inside the funnel at $r=6.3$. Through shock discontinuity, the jet jumps from supersonic branch to subsonic branch and then again accelerates and becomes transonic at $r_c=8.62$ reaching at terminal speed $v_t=0.86$. The origin of generation of shock as well as multiple sonic points lies in radiation drag and negative flux inside the funnel (Fig. \ref{lab:rad_mom}b, blue dotted) which collectively resist the jet and it forms multiple sonic points. The details of calculation of shock conditions are described in \citep{vc18a,vc18b} and are not repeated here.
 
\subsection{B-type solutions : Hot and slow jet base}
\begin {figure}
\begin{center}
 \includegraphics[width=9cm, trim=0 0 0 300,clip]{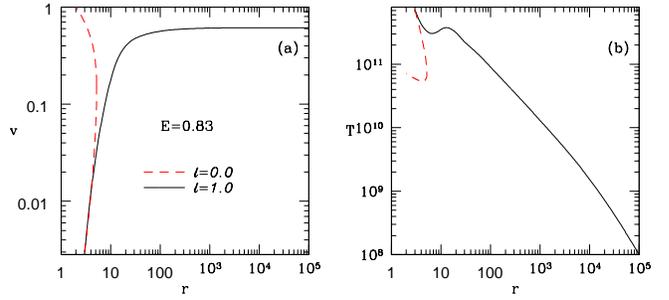}
\vskip -0.5cm
 \caption{Nature of B-type solutions. Variation of (a) $v$ and (b) $T$ with $r$ for $E=0.83$ and $\ell=1.0$ (solid, black) and $\ell=0$ (dashed, red).}
\label{lab:B_type}
 \end{center}
\end{figure}
\begin {figure}
\begin{center}
 \includegraphics[width=7cm, trim=0 0 0 7cm,clip]{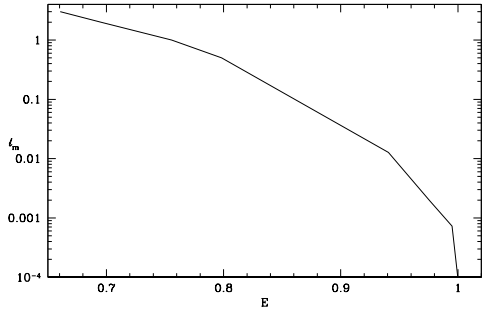}
\vskip -0.5cm
 \caption{Minimum $\ell$ i.e., $\ell_{\rm m}$ as a function of $E$ for B-type solutions.}
\label{lab:minl}
 \end{center}
\end{figure}

Class B are bound solutions close to the horizon i.e., $E<1$ extending up to infinity driven by radiation field (Fig. \ref{lab:gen_sol}, solid, black). These solutions are characterized by low base speeds but high base temperature. In other words, the kinetic component of $E_b$ is ineffective ($\gamma_b\sim 1$). The thermal component is high ($h_b>1$) but dominated by gravity $E\sim h_b \sqrt{g^{rr}_b}<1$. Hence the thermal driving by itself is unable to push the matter outward (dashed, red) and folds back onto the horizon,
while the radiatively driven flow is transonic and escapes the gravity of the black hole. The matter is pushed outward {collectively by radiation momentum deposition and energy transfer} onto the jet. As these solutions are absent in elastic scattering regime, Compton scattering is essential for driving jets with $E<1$. We choose $E=0.83$ and plot $v$ and $T$ with $r$ in Figs. (\ref{lab:B_type}a and \ref{lab:B_type}b) respectively for the choice of $\ell=1$ (solid, black) and compare the same with thermally driven flow (i. e., flow with $\ell=0$, dashed, red). At $r_b=3$, $v_b=0.003$ and $T_b\sim7 \times 10^{11} \rm K$ for the both the flows, the radiatively driven jet (solid, black) flows to infinity through a sonic point $r_c=13.54$ and achieves a terminal speed $v_t\sim 0.61$, but the thermally driven flow
cannot expand to infinity against the gravity of the central black hole. This brings us to the question
what is the minimum disc luminosity required to blow a jet starting with specific energy $E<1$.  For $E>1$
(i.e., A type) jets, a transonic solution is guaranteed even for $\ell=0$. As has been shown in Figs. (\ref{lab:B_type}a, \ref{lab:B_type}b) for B-type flows, it only blows as jet if and only if $\ell>0$. In Fig.
(\ref{lab:minl}) we plot minimum disc luminosity $\ell_{\rm m}$ required to blow a transonic jet for B-type jets i.e., as
a function of $E$. For example, Fig. \ref{lab:minl} shows, we can have transonic jet for $E<0.75$ if the disc luminosity is $\ell>1.0$.

\subsection{C-type solutions : Cold and very slow jet base}
\begin {figure}
\begin{center}
 \includegraphics[width=9cm, trim=0 0 0 10cm,clip]{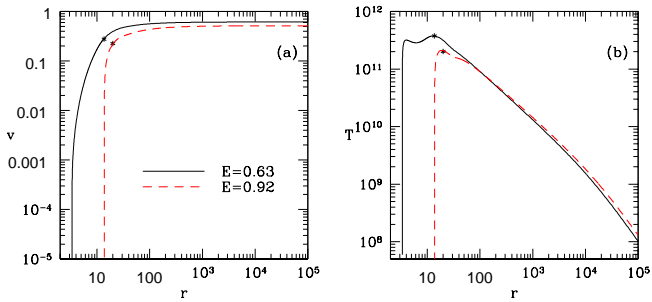}
\vskip -0.5cm
 \caption{Nature of C-type solutions. Variation of (b) $v$ and (c) $T$ with $r$ for $E=0.63$ (solid, black) and
 $E=0.92$ (dashed, red) for $\ell=1.0$}
\label{lab:C_type}
 \end{center}
\end{figure}

\begin {figure}
\begin{center}
 \includegraphics[width=9cm, trim=0 0 0 1cm,clip]{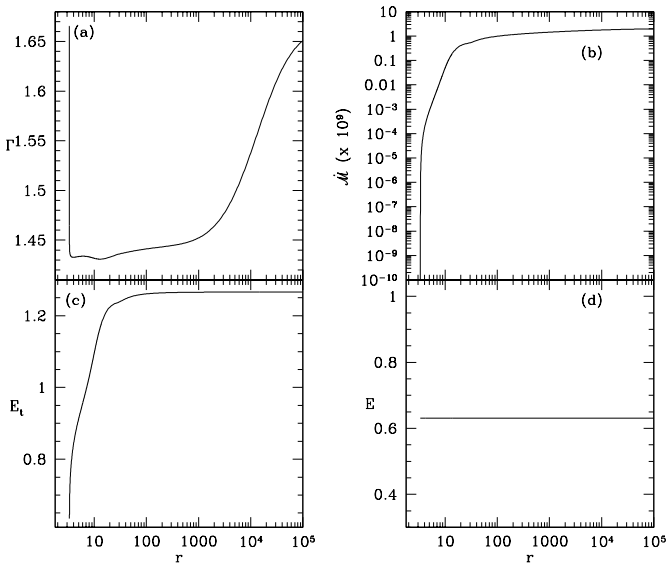}
\vskip -0.5cm
 \caption{Variation of (a) $\Gamma$ and (b) $\mdtj$, (c) $E_t$ and (d) $E$ for $E=0.63$ (solid) and $\ell=1.0$}
\label{lab:gam_poly}
 \end{center}
\end{figure}

Solutions corresponding to red dashed line in $E-r_c$ plot (Fig. \ref{lab:gen_sol}) attract special attention, as the jets represented by these solutions are characterized by $E<1$ but additionally, the bases of the jet
are different and $r_b > 3$. So for a given $r_b$ and $E$, there is a maximum $\ell$ for which $v_b \rightarrow 0$. So one can find jet solutions with the same $E$ and $\ell$, if $r_b$ is increased and these later solutions
are so-called C-type. Typically, the thermal ($T_b<<10^{10}$) and kinetic ($v_b\sim 0$) components of $E$ at the jet base is very low. 
In Fig. (\ref{lab:C_type}a-b), we plot $v$ and $T$ with $r$ for $E=0.63$ (solid, black)
and $E=0.92$ (dashed, red) and powered by disc radiation of $\ell=1.0$.
Both the jets are characterized by single sonic points (star marks)
and the terminal speeds attained by both jets are around $v_t \sim 0.6$. The base temperature and three velocity
are non-relativistic for both the jets. The temperature profile of both the jets
rises from $T_b\sim 10^7 \rm K$ values to $3.7\times10^{11}\rm K$ (solid, black) and $2\times 10^{11}{\rm K}$ (dashed, red) because of Compton heating. The base of the $E=0.63$ jet is at $r_b\gsim 3$, but for $E=0.92$
it is $r_b=12$ i.e., quite different, but they generate
somewhat similar terminal quantities. The mechanism of the heating and cooling can be understood if we look at equation (\ref{dthdr.eq}). First three terms inside the square bracket are positive and responsible for the decrease in jet temperature due to expansion.
The last term in the bracket shows radiative heating. For $\ep$ flow, the heating term can be written as :
\be
Q^+ \approx (\Gamma-1)(1-\chi_c) \left[\frac{g^{rr} {\cal R}_0}{v}+\frac{v {\cal R}_2}{g^{rr}}-2{\cal R}_1\right]
\label{therm_acc.eq}
\ee
At the base $\Theta_b \approx v_b \approx$ small, moreover, ${\cal R}_1<0$ inside the funnel, i.e., all the terms inside the bracket are positive and collectively heat up the jet near the base. In addition, since $v_b \sim$small, then the first term within the parenthesis of equation (\ref{therm_acc.eq}) is dominant and is responsible for the sharp rise in temperature at $r\sim r_b$. 
Interestingly, there is a second hump in $T$ profile of the jet. Within one Schwarzschild radius from the jet-base,
$v$ increases by more than four orders of magnitude ($\sim 10^{-5}~\rightarrow 0.1$), so the first term tends to decrease while second term
$v{\cal R}_2/g^{rr}$ starts to become important. The shape of the temperature profile is influenced by the relative strength of these two terms { within first} few Schwarzshild radii from the base. If one carefully studies the distribution of radiative moments (Figs. \ref{lab:rad_mom}a, b), then around $10< r < 20$ the flux becomes positive $R_1>0$, { beyond which,} the third term $2R_1$
starts to dominate over both the first and the second {terms} in the parenthesis of equation (\ref{therm_acc.eq}). This
leads to decrease in temperature and eventually the second peak. So the first hump is due to the first term in the parenthesis and the second due to the interplay of all the the three terms especially second and the third. Although for larger $r$, Compton heating is not important and the jet cools down due to expansion.

In Fig. (\ref{lab:gam_poly}a-b) we plot variation of $\Gamma$ and $\mdtj$ corresponding to parameters of the jet corresponding to $E=0.63$ of the Fig. (\ref{lab:C_type}). Variation of $\Gamma$ delivers similar information that plasma is cold and non-relativistic at the base as well as far away ($r\sim10^5$), but radiation makes it relativistic and hot in between. Variation of Entropy depicts non-adiabatic nature of the jet as $\mdtj$ increases by around $10$ orders of magnitude.
As expected, $E_t$ (Fig. \ref{lab:gam_poly}c) evolves and increases due to the impact of radiation. Starting from $E_{tb}=0.63<1$ at the base, it reaches at $E_t \sim1.206>1$ while the generalized relativistic Bernoulli parameter $E$ remains conserved and is shown to be a constant of motion (Fig. \ref{lab:gam_poly}d).

\subsubsection*{Effect of luminosity on C-type jet}
\begin {figure}
\begin{center}
 \includegraphics[width=8.5cm, trim=0 0 0 0,clip]{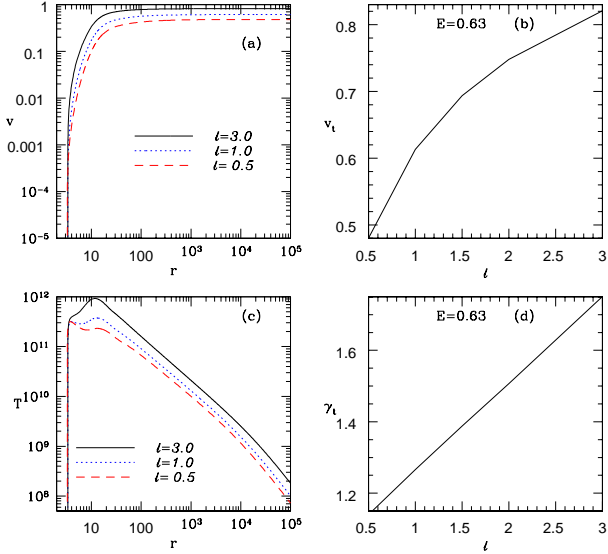}
\vskip -0.5cm
 \caption{(a) Velocity profiles for various luminosities for C-type solutions for $\ep$ composition. Corresponding terminal speeds ($v_t$) are plotted in (b). (c) Variation of $T$ with $r$ for various luminosities. (d) Lorentz factor ($\gamma_t$ ) as a function of $\ell$ (solid black) In panels (a) and (c) different curves are for $\ell=3.0$ (solid black), $\ell=1.0$ (dotted blue) and $\ell=0.5$ (dashed red) keeping $E=0.63$.}
\label{lab:Temp_profile}
 \end{center}
\end{figure}

We keep same $E=0.63$ and plot velocity profiles for $\ell=3.0$ (solid black), $\ell=1.0$ (dotted blue) and $\ell=0.5$ (dashed red) in Fig. (\ref{lab:Temp_profile}a). As expected, greater acceleration is observed as the radiation field gets more intense. To estimate qualitative magnitude of acceleration and effect of $\ell$, we plot $v_T$ with $\ell$ for $E=0.63$ in Fig (\ref{lab:Temp_profile}b). The terminal speeds range from $v_t=0.53$ to $v_t=0.82$ as $\ell$ goes from $0.5$ to $3.0$.\\
The corresponding temperature profiles for these luminosities are shown in Fig. (\ref{lab:Temp_profile}c).
Out of the two peaks, the first one is weakly dependent on $\ell$, because it is mostly dictated by $v_b \rightarrow$small (i. e., the first term in the parenthesis of r.h.s of equation \ref{therm_acc.eq}). While the second one depends on $\ell$, because the second peak is borne by the combined effect of all the moments. 

\subsection{Effect of composition on jet dynamics}
\begin {figure}
\begin{center}
 \includegraphics[width=8.5cm, trim=0 0 0 0,clip]{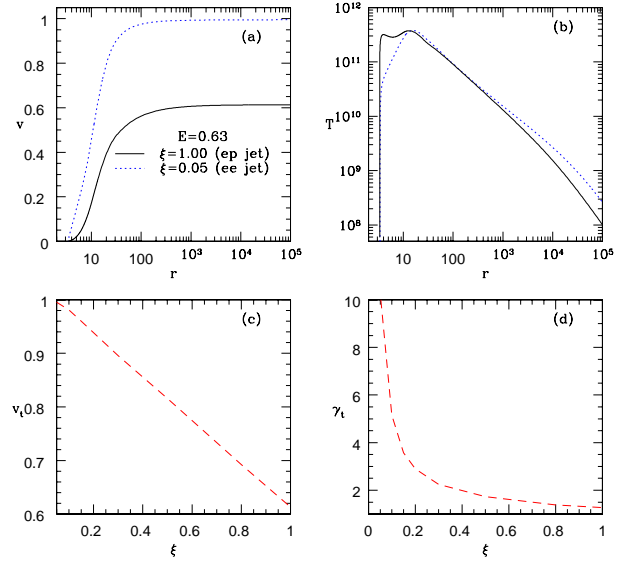}
\vskip -0.5cm
 \caption{(a) $v$ and (b) $T$ profiles as a function of $r$ for $\xi=0.05$ (dotted blue) and $\xi=1.00$ (solid black). 
 (c) $v_t$ and (d) $\gamma_t$ as a function of $\xi$.
For all the curves $E=0.63$ and $\ell=1.0$.}
\label{lab:Temp_profile_xi}
 \end{center}
\end{figure}
Composition of the relativistic jets is a much debated topic. Jets are believed to be dominated either by baryons ($\ep$ plasma) or by leptons (dominated by $\el$). { As we have considered relativistic EoS, which} takes care of composition of the plasma through $\xi$ {(Appendix \ref{app:EoS}) }. It permits us to study the jet dynamics with variation of $\xi$. 
To study the effect of composition, we generate solutions by varying $\xi$ for given values of $E$ and $\ell$.  We plot three velocity $v$ in Fig. (\ref{lab:Temp_profile_xi}a) and temperature $T$ in Fig. (\ref{lab:Temp_profile_xi}b) for $\xi=1.0$ (solid black) and $\xi=0.05$ (dotted blue) for $E=0.63$ and $\ell=1.0$. As $\xi$ decreases, lepton fraction in fluid composition increases making the fluid lighter, hence the jet under radiative acceleration, becomes faster. Corresponding terminal speeds are plotted in Fig. (\ref{lab:Temp_profile_xi}c) which go up to $0.998$ as the value of $\xi$ drops to $0.05$. In terms of terminal Lorentz factors $\gamma_t$ of the jets (Fig.\ref{lab:Temp_profile_xi}d), for very low $\xi(\sim 0.05)$, $\gamma_t$ comfortably reaches up to $10$. The temperature profiles of a baryon dominated jet (solid, black) is mostly similar to
lepton dominated jets (dotted, blue). For thermally driven jets or jets driven by radiation in the elastic scattering regime, the temperature of baryon dominated flows is greater than lepton dominated ones, however,
in the Compton scattering regime, the energy transferred by the radiation is more effective for flows with lower $\xi$ (presence of $\rho_e$ in r.h.s of equation \ref{en1con.eq}).

\begin {figure}
\begin{center}
 \includegraphics[width=8.5cm, trim=0 0 0 3cm,clip]{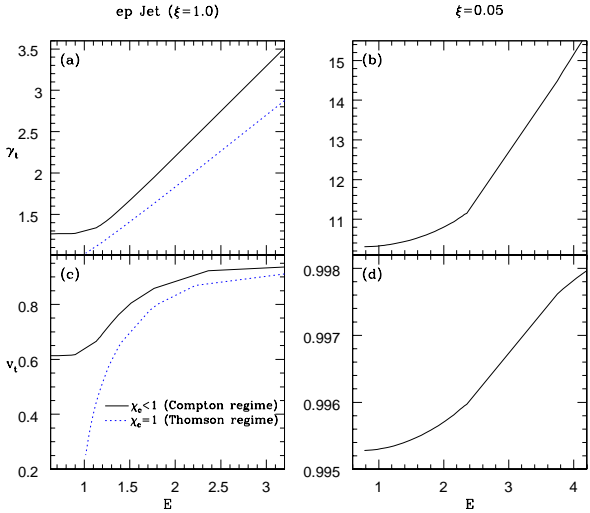}
\vskip -0.5cm
 \caption{Terminal Lorentz factor $\gamma_t$ (a, b) and terminal speed $v_t$ (c, d) as a function of $E$.
The composition of jets are $\xi=1$ (a, c) and $\xi=0.05$ (b, d). 
The quantities in the Compton scattering (solid, black) is compared with Thomson scattering regime (dotted, blue) in panels (a) and (c). All the plots are for $\ell=1.0$.}
\label{lab:term_E_xi}
 \end{center}
\end{figure}

We plot $\gamma_t$ and $v_t$ with $E$ for $\xi=1.0$ (Fig. \ref{lab:term_E_xi} a,c) and $
\xi=0.05$ (dashed, blue; Fig. \ref{lab:term_E_xi} b,d). The curves are plotted for $\ell=1.0$. For $\ep$ jets we compare terminal quantities of current 
paper (solid, black) with those obtained in the Thomson scattering regime \citep[dotted, blue, similar to][]{vc18b}, which reasserts the fact 
that Compton scattering accelerates the jets more effectively than in the Thomson regime. Further, as there are no 
solutions for $E<1$ in the Thomson scattering regime, the terminal speeds approach very low values as
$E\rightarrow 1$. However, Compton driven jets maintain $v_t>0.6$ even for $E<1$. This lower limit off $v_t$ or
$\gamma_t$, is highly relativistic for ($v_t>0.99$) for lepton dominated jets (dashed, blue).
It may be noted Figs. (\ref{lab:Temp_profile_xi}a-d) is for C-type jets, but Figs. (\ref{lab:term_E_xi})
represents all types of jets for given jet and radiative parameters.
 
\section{Discussion and concluding remarks}
\label{sec:con}
In this paper, we { have }studied radiatively driven relativistic jets in the Compton scattering regime and curved space-time.
The thermodynamics of the jet is described by relativistic EoS.
This work is in continuation of our previous efforts where we studied interaction between radiation and jet
matter in the elastic scattering regime. In this paper, we show that the radiative driving in the Compton regime
is significantly more effective than in the elastic scattering regime. 
In the Compton regime, both energy and momentum are transferred to the jet from the radiation field, so the radiation not only accelerates the flow, but {also} increases its temperature. Consequently, we showed three classes or types of jets and named them A, B, and C. Class A jets were those which {are launched} with very high temperatures and high speeds at the base with $E>1$,
class B are those which start with relatively less speeds and temperatures. And class C are the ones,
{that have very low base temperatures} and very small base speeds. Both B \& C type jets have $E<1$.
This is because radiation transfers
energy to the matter so that flows with $E<1$ can also be ejected.
In scattering regime, at such $E$, the jet matter ejected outwards would actually fall back to the BH. The velocity distribution would almost overlap
with the thermal one (dashed, red in Fig. \ref{lab:B_type}a). For $E<1$, the terminal Lorentz factors obtained for super Eddington luminosities are below 2 for $\ep$ jets but they are sufficiently high and reach beyond 10 for lepton dominated jets. We reiterate that our previous papers \citep{vkmc15,vc18a,vc18b} had no counterpart 
of the B and C type jets, because in the Thomson scattering regime a transonic jet always has $E>1$.

Astrophysical jets are transonic, fast and collimated and hence assumption of conical or spherical jets is very common in theoretical investigation. If the jet was assumed to be adiabatic then assumption of
spherical jet is generally a fair one. However, in the Compton regime radiation transfers energy to the jet, thereby heating it. So there are grounds for apprehension whether the lateral expansion of the jet could destroy the spherical symmetry. In Appendix \ref{cross_section}, we showed that for transonic jets, the lateral
expansion is much smaller than the radial expansion. In Fig. \ref{lab:figC2} we plotted ${\cal K}$ as a function of $r$ where, ${\cal K}={\rm a}_r/{\rm a}_x$ or the ratio between radial acceleration and the lateral
pressure gradient term. It is clear from the figure that ${\cal K}>$few, even in subsonic region.
 
Radiatively driven jets possess multiple sonic points and internal shocks in certain range of the parameters. These internal shocks may be produced due to various factors. We showed in \citet{vc17} that non radial cross section may harbour internal shocks. In \citet{vc18a,vc18b}, we showed that in presence of radiation, even radial jets may undergo shock transitions. The shocks obtained in this paper have similar features.
A number of processes that give rise to internal shocks give theoretical support and strengthen the attempts that assumed internal shocks to explain various observed features of the jets \citep{bk79, l11}, mainly the high energy flux (GeV to TeV) of the radiation spectrum. 
 
In this paper, we have obtained jets with a variety of terminal speeds ranging from mildly relativistic
to highly relativistic. While {jets in microquasars do show a range of terminal speeds,
it} is assumed that the astrophysical jets are relativistic.
Bulk speeds of the jets are inferred from crude observational methods. 
However, the presence of asymmetry in luminosity of a jet and its counter part puts better
constraint on the estimation of Lorentz factors of the jets \citep{wa97,hk06}. The range of terminal Lorentz factors obtained in this paper is similar to the ranges obtained for Lorentz factors in both X-ray binaries as well as AGNs \citep{mjfn06}.  

Quantitatively, we can conclude that choosing a jet base at $r_b=3$, $\ep$ jets can be accelerated to a minimum of
of $v_t \sim 0.22$ for disc luminosity $\ell=0.1$. For the same $r_b$ and $\ell$ 
of lepton dominated jets ($\xi=0.1$), the minimum terminal speed is $v_t \sim 0.78$. Of course for
luminous discs ($\ell=1$), terminal speeds for hot $\ep$ jets can go up to $v_t>0.9$ while
for lepton dominated jets terminal Lorentz factors are ultra relativistic. 
\appendix
\section{Relativistic equation of state}
\label{app:EoS}
At relativistic temperatures, the adiabatic index ($\Gamma$) depends upon temperature. The value of $\Gamma$ ranges from $5/3$ to $4/3$ as the flow goes from non relativistic temperatures to relativistic temperatures.
We consider EoS for multi species, relativistic
flow proposed by \citet{c08,cr09}, which is a close approximation
of the exact relativistic EoS \citep{c38,s57,vkmc15}. The EoS is 
given as,
\begin{equation}
e=n_{e^-}m_ec^2f, \mbox{ in physical dimensions}
\label{eos.eq}
\end{equation}
where, $m_e$ and $n_{e^-}$ are the rest mass of the electron and electron number density respectively. $f$ is a dimensionless quantity
given by
\begin{equation}
f=(2-\xi)\left[1+\Theta\left(\frac{9\Theta+3}{3\Theta+2}\right)\right]
+\xi\left[\frac{1}{\eta}+\Theta\left(\frac{9\Theta+3/\eta}{3\Theta+2/\eta}
\right)\right].
\label{eos2.eq}
\end{equation}
Here,
$\Theta=kT/(m_ec^2)$ is dimensionless temperature ($T$) and $k$ is Boltzmann constant.
$\xi (= n_{p^{+}}/n_{e^{-}})$ is ratio of number densities of 
protons and electrons.
$\eta (= m_{e}/ m_{p^{+}}$) is mass ratio of electron and proton.
The expressions of $\Gamma$, $N$,
$a$ and $h$ (in geometric units) are given by
\begin{equation}
N=\frac{1}{2}\frac{df}{d\Theta} ;~~ \Gamma=1+\frac{1}{N} ; ~~
a^2=\frac{\Gamma p}{e+p}=\frac{2 \Gamma \Theta}
{f+2\Theta}.; ~~ h=\frac{f+2 \Theta}{\tau}
\label{sound.eq}
\end{equation}
Here $\tau$ is a function of composition and is defined as $\tau=2-\xi+\xi/\eta$.
\section{Radiation field from accretion disc and associated radiation parameters}
\label{app:rad_mom}
\subsection{Estimating approximate accretion disc variables}
We have $U^\mu$ being
four-velocity components in the accretion disc and $v\equiv(\vartheta_x,0,\vartheta_\phi)$ are corresponding three-velocity components with $x,~\theta,~\phi$ are spatial coordinates. $\vartheta=\vartheta_x/\sqrt{(1-\vartheta_\phi^2)}$ is defined as the radial component of three-velocity measured
by local observer rotating along the disc surface. We can show
the velocity distribution of the outer disc and the corona in following compact form
[see Appendix A of VKMC15]
\be
\vartheta_{\rm i}=\left[1-\frac{(x-2)x^2}{\{x^3-[(x-2)\lambda^2]\}U_t^2|_{x_{0\rm i}}}
\right]^{1/2}.
\label{accvel.eq}
\ee
Here, the suffix ${\rm i}$ represents variables associated with the corona (i. e.,
i$=${\small C}) and the outer disc (i. e., i$=${\small D}).
$U_t|_{x_{0\rm i}}$ is the covariant time component of the $U^\mu$s at the outer boundary of the respective disc component, which are $x_{0\rm i}=\xsh$ and $x_{0\rm i}=x_0$ for the corona and the outer disc respectively.
At $x_0$, $ [\vartheta_{\rm \small D}]_{x_0} \approx 0$ and it grows as it falls inwards.

The distribution of temperature can be obtained as \citep{vkmc15}
\be
\Theta_{\rm i}=\Theta_0\left(\frac{U^x_0x_{0 \rm i}H_0}{U^x_{\rm i} xH_{\rm i}}\right)^{\Gamma -1}.
\label{acctemp.eq}
\ee
The compression ratio $R_c$ is defined as 
\be 
R_c=U^r_-/U^r_+
\ee
where $R_c$ is approximated from the results of \cite{kc17},
\bea
R_c=2.46-2.12\times 10^{-3} \xsh-4.72\times 10^{-4} \xsh^2 \nonumber \\
+5.98\times 10^{-6} \xsh^3-2.08\times 10^{-8} \xsh^4
\eea

Similarly the temperature increases by the same fraction $R_c$ giving outer boundary velocity and temperature for Corona, using which in equations (\ref{accvel.eq}) and (\ref{acctemp.eq}), $\vartheta_{\rm C}$ and $\Theta_{\rm C}$ are obtained.

Moreover, \citet{vkmc15} proposed a relation between $\xsh$ and accretion rate ${\dot m}$, which, after converting into current unit system, becomes
\be
\xsh=87.402-28.193{\dot m}+3.125{\dot m}^2-0.115{\dot m}^3
\label{xsdotm.eq}
\ee
$\xsh$ is in geometric units while $\dot m$ is mass accretion rate in Eddington units
(Eddington accretion rate is defined as $\equiv {\dot M}_{\rm Edd}=1.4\times10^{17}\mbh/\msol$gs$^{-1}$).
To specify $\vartheta_{\rm i}$ and $\Theta_{\rm i}$ at $x$, the local height $H_{\rm i}$ is also required.
Following numerical simulations \citep{dcnm14,lckhr16}, we define $H_{\rm 0}=0.4H_{\rm sh}+\tan \theta_{\rm D}x_0$.
Supplying values of all parameters required, $[\vartheta_{\rm \small D}]_{x_0},~\rho_0,~H_0$ and ${\dot m_\sk}$ at outer disc boundary, $x_0$, velocity, temperature and density at all $x_{\rm i}$ along with the location of $\xsh$ is obtained. The
accretion parameters considered in this paper
are shown in table \ref{table2}.
\begin{table}
\caption{Disc parameters}
\label{table2}
\centering
 \begin{tabular}{|c c c c c c c|} 
 \hline
 \hline
 $\lambda$ & $x_0$ & $\left[\vartheta_{\rm \small D}\right]_{x_0}$ & $\left[\Theta_{\rm \small D}\right]_{x_0}$ & $\theta_D$ & $H^*$ & $d_0$\\ [0.5ex] 
 \hline
 $3.6$ & $20000 \rs$ & $0.001$ & $0.03$ & $78.5^0$ & $40$ & $0.4H_{\rm sh}$\\ 
 \hline
 \end{tabular}
\end{table}
\subsection{Radiative intensity and luminosity from the accretion flow}
Assuming stochastic magnetic field in the accretion disc and considering that it is in partial equipartition with gas pressure, we can assume that the ratio between magnetic pressure ($p_{\rm mag}$)
and the gas pressure ($p_{\rm gas}$) is constant $\beta$
{\ie} $p_{\rm mag}=B^2/8\pi=\beta p_{\rm gas}=\beta n k T$.
The outer disc emits through synchrotron and bremsstrahlung processes while the corona additionally emits 
through inverse-Compton process along with these. 
The frequency integrated, local intensity for outer disc is \citep{kc14,vkmc15}, 
$$
{\tilde I}_{\rm i_0}={\tilde I}_{\rm syn}+{\tilde I}_{\rm brem}
$$
\bea
=\left[\frac{16}{3}\frac{e^2}{c}\left( \frac{eB_{\rm i}}{m_e c} \right)^2
 \Theta^2_{\rm i} n_{\rm i}+ 1.4\times 10^{-27}n_{\rm i}^2g_bc \sqrt{\frac{\Theta_{\rm i} 
m_e}{k}}\right] \nonumber \\
\times \frac{\left(d_0 \sin \theta_{\rm i}+x\cos \theta_{\rm i}
\right)}{3} \ \  {\rm erg} \
{\rm cm}^{-2} {\rm s}^{-1}
\label{skint.eq}
\eea
Here, $x, n_{\rm i}, \Theta_{\rm i}, \theta_{\rm i}, g_b(=1+1.78\Theta_{\rm i}^{1.34})$ and $B_{\rm i}$ are the 
radial distance, electron number density, local dimensionless temperature, the semi-vertical
angle, 
relativistic Gaunt factor and the magnetic field respectively for both disc components. 
The factor multiplied outside the square bracket
converts the emissivity (${\rm erg}~{\rm cm}^{-3}{s}^{-1}$) into intensity  (${\rm erg}~ {\rm cm}^{-2}{s}^{-1}$).
Now, the emitted radiation from the outer disc is inverse Comptonized within the disc, adding which, the specific intensity becomes \citep{pk95, by76},
\be
{\tilde I}_{\rm D}={\tilde I}_{\rm D_0}\left(1+4\Theta_{De}+16\Theta_{De}^2\right)^{\tau_r+\frac{\tau_r^2}{3}}
\label{disc_int.eq}
\ee
Here $\Theta_{De}$ is dimensionless electron temperature of the outer disc component and $\tau_r$ is optical depth, which depends upon $r$ and obtained to be,
\be
\tau_r=\frac{K_0 \dot{m}}{u^rrh}
\ee
Where 
$$
K_0=\frac{1.44\times10^{17} \sigma_Tc}{4m_e(1+1/\eta)G\pi M_{\odot}}
$$
The outer disc luminosity is obtained by integrating ${\tilde{I}}_{D}$ over the whole disc surface, {\ie}
\be
L_{\od}=2\int^{x_{0}}_{\xsh} \int^{2\pi}_{0} {\tilde{I}}_{\od}r\left(1-\frac{2}{r}\right)^2~{\rm cosec}^2\theta_{\od}~d\phi dx
\label{lum.eq}
\ee

Now, a fraction of radiation emitted from outer disc falls onto the corona and adds to the radiation emitted by it and is calculated to be,

\bea
L^f_{\od}=2\int^{x_{0}}_{\xsh} \int^{2\pi}_{0} {\tilde{I}}_{\od}r\left(1-\frac{2}{r}\right)^2\frac{1}{\pi}\tan^{-1}\left(\frac{\xsh}{x}\right) \tan^{-1}\left(\frac{H_{\rm sh}}{x}\right)\nonumber \\
\times~{\rm cosec}^2\theta_{\od}~d\phi dx
\eea

Assuming that this radiation falls on to the corona homogeneously, we can calculate the additional specific intensity of the corona which takes this radiation into account as,

\be 
{\tilde I}_{\rm C^f_0}=L^f_{\od}/{A_{\rm \small C}}
\label{corona_int.eq}
\ee
Including this radiation along with local inverse Comptonization inside corona, the coronal luminosity is obtained as

\bea
L_{\rm C}=2\int^{\xsh}_{x_{\rm ii}} \int^{2\pi}_{0} \left[{\tilde I}_{\rm C_0}+{\tilde I}_{\rm C^f_0}\right]\left(1+4\Theta_{\rm Ce}+16\Theta_{\rm Ce}^2\right)^{\tau_r+\frac{\tau_r^2}{3}} \nonumber \\
\times ~~~r\left(1-\frac{2}{r}\right)^2~{\rm cosec}^2\theta_{\rm C}~d\phi dx 
\eea
Similarly, here $\Theta_{\rm Ce}$ is dimensionless electron temperature of the corona.\\ 
Above luminosities can be presented in units of $L_{\rm Edd}(\equiv 1.38\times10^{38}\mbh/\msol~{\rm erg~}{\rm s}^{-1})$ as $\ell_{\rm i}=L_{\rm i}/L_{\rm Edd}$.\\
Considering corona to be compact and having isotropic distribution of radiation, we can have relation of specific intensity of the corona in terms of $L_{\rm C}$ as

\be 
{\tilde I}_{\rm \small C}=L_{\rm \small C}/{\pi}{A_{\rm \small C}}={\ell_{\rm \small C}}L_{\rm Edd}/{\pi}{A_{\rm \small C}}~({\rm erg}~ {\rm cm}^{-2} {\rm s}^{-1}),
\ee
\subsection{Radiative moments}
\subsubsection{Relativistic transformation of intensities from various disc components}
To solve the EoMs of the jet, we require information of radiation field, governed by radiative moments on the jet axis. To get radiative moments, we need to compute specific intensities from the outer disc as well as corona. Using the expressions of velocity (\ref{accvel.eq}) and temperature (\ref{acctemp.eq}) from both the disc components we compute the radiative intensity (\ref{skint.eq}, \ref{disc_int.eq}, \ref{corona_int.eq}) in the local rest frame of the disc, which is transformed into the curved frame, following special and general relativistic transformations as
\begin{equation}
I_{\rm i}=\frac{{\tilde I}_{\rm i}}{\gamma^4_{\rm i}\left[1+{\vartheta}_jl^j\right]^4_{\rm i}}\left(1-\frac{2}{x}\right)^2
\label{Itrans.eq}
,\end{equation}
where ${\tilde I}_{\rm i}$ is frequency integrated specific intensity in local rest frame of the disc, ${\vartheta}^j$ is $j^{\rm th}$ component of three-velocity,  $\gamma_{\rm i}$ being the Lorentz factor, $l^j$s are direction cosines. 
The square of redshift factor $(1-2/x)^2$ reduces radiation intensity close to the BH \citep{b02}.

\subsubsection{Calculation of radiative moments in curved space-time}
Zeroth, first, and second moments of specific intensity,
(\ie $\int I d\Omega;\newline \int I l^j d\Omega;~\ $ $\&$ $~ \int I l^j l^k d\Omega$, respectively) contain all the information of radiation field. These are
ten independent components \citep{mm84,c05}. But to study a conical narrow jet, passing through radiation field, only three of these are dynamically important.   

The radiative moments ($R_0, ~ R_1~\&~R_2$) can be written in a compact form given by \citep{vc18a},
\bea
R_{n\rm i}=\int^{x_{\rm i 0}}_{x_{\rm ii}} \int^{2\pi}_{0}\left(1-\frac{2}{x}\right)^3\frac{{\tilde I}_{\rm i}}{\gamma^4_{\rm i}\left[1+{\rm v}_jl^j\right]^4_{\rm i}} \nonumber \\
\times \left[\frac{(r-x \cos\theta_{\rm i})}{\sqrt{[(r-x \cos\theta_{\rm i})^2+x^2\sin\theta_{\rm i}^2]}}\left(1-\frac{2}{x}\right)+\frac{2}{x}\right]^n \nonumber \\
\times \frac{rxd\phi dx}{[(r-x \cos\theta_{\rm i})^2+x^2\sin\theta_{\rm i}^2]^{3/2}},
\label{moments2.eq}
\eea
Here ${\rm i} \rightarrow {\rm C}~\ $ and $~ {\rm D}$ signifies the contribution from the corona and the
outer disc, respectively. The integration is performed over $x$ and $\phi$ with the limits of integration being from $x_{\rm ii}$  (inner edge) to $x_{\rm i 0}$ (outer edge) of the  respective disc component and the angular circumference of the disc from $0$ to $2 \pi$.
The index $n=0, 1, 2$ is for
$R_0,~R_1~\&~R_2$, which are, radiative energy density, radiative flux along $r$ and the $rr$ component
of the radiative pressure respectively. $\theta$ is semi vertical angle of the respective disc component (Fig. \ref{lab:fig1}). 

$\gamma^4_{\rm i}\left[1+{\rm v}_jl^j\right]^4_{\rm i}$ in the denominator inside integration represents special relativistic transformation of specific intensity \citep{c05} while $2/x$ accounts for transformation of ${\tilde I}_{\rm i}$, $l^j$ and solid angle in curved space-time. These transformations are taken from methods developed by \cite{b02, bgjs15} and were used by \cite{vc18a}.
 
As we have two disc components, corona and outer disc, the total moments are obtained as,
\be 
R_n=R_{n{\rm \small C}}+R_{n{\rm \small D}}
\label{moments3.eq}
\ee

The $x$ limits for corona are
$x_{\rm {\small C}i}=2, x_{{\rm \small C} 0}=\xsh$. Following the shading effect induced by corona as it blocks certain amount of radiation from the outer disc, the innermost edge seen from $r$ is given by,

$$
x_{\rm {\small D}i}=\frac{r-d_0}{(r-H_{\rm sh})/\xsh+ \cot \theta_{\rm \small C}}.
$$
It is clear from above that as $r\rightarrow \infty$,
$x_{\rm {\small D}i}\rightarrow \xsh$. Further, up to some $r=r_{\rm lim}$, radiation from the outer disc won't
reach the jet axis. This limiting distance is obtained as, 

\be
r_{\rm lim}=\frac{x_0H_{\rm sh}-H_0\xsh}{x_0-\xsh}.
\label{shadolim.eq}
\ee

\section{On validity of the spherical cross section of the jet}
\label{cross_section}
As the radiation transfers energy to the jet matter, the jet heats up increasing the pressure of the flow. Pressure gradient force being isotropic in the local frame might compete with the radial expansion of the flow and the conical flow geometry of the jet may be compromised.

To justify this approximation, here we show that the outward acceleration of the jet is much more dominant over lateral expansion.

The jet acceleration along $r$ is given by ${\rm a}_r$ (right hand side of equation Eq. \ref{eu1con.eq}),
\begin{equation}
{\rm a}_r=-\left(1-\frac{2}{r}+u^ru^r\right)
\frac{1}{e+p}\frac{dp}{dr}+{\rho}_e\frac{{\sqrt{g^{rr}}\gamma^3}
}{(e+p)}{\Im}^r
\end{equation}

Let $x=r {\rm sin}\alpha$ be the horizontal radius of the jet boundary and $\alpha$ is the opening half angle.
Since $\alpha$ is small, therefore
$$
\frac{dp}{dx} \approx \frac{dp}{dr}\frac{\delta r}{\delta x}
$$

Using the above expression, the horizontal pressure gradient term can be written as

\be
\frac{dp}{dx}\sim\frac{dp}{dr}\frac{\delta r}{\delta x}=\frac{(1-{\rm cos}\alpha)}{{\rm sin} \alpha}\frac{dp}{dr}
\ee
This gradient of pressure  from jet axis to the jet wall leads to thermal expansion of the jet. Hence there is a net acceleration component of the jet away from jet axis (be it ${\rm a}_x$), which can be written as

\be
{\rm a}_x = -\left(1-\frac{2}{r}\right)\frac{1}{e+p}\frac{dp}{dx}=-\left(1-\frac{2}{r}\right)\frac{1}{e+p}\frac{(1-{\rm cos}\alpha)}{{\rm sin} \alpha}\frac{dp}{dr}
\ee

Now to compare ${\rm a}_r$ and ${\rm a}_x$, we define
\be 
{\cal K}=\frac{{\rm a}_r}{{\rm a}_x}
\ee

\begin {figure}
\begin{center}
 \includegraphics[width=9.cm, trim=0 0 3cm 3cm]{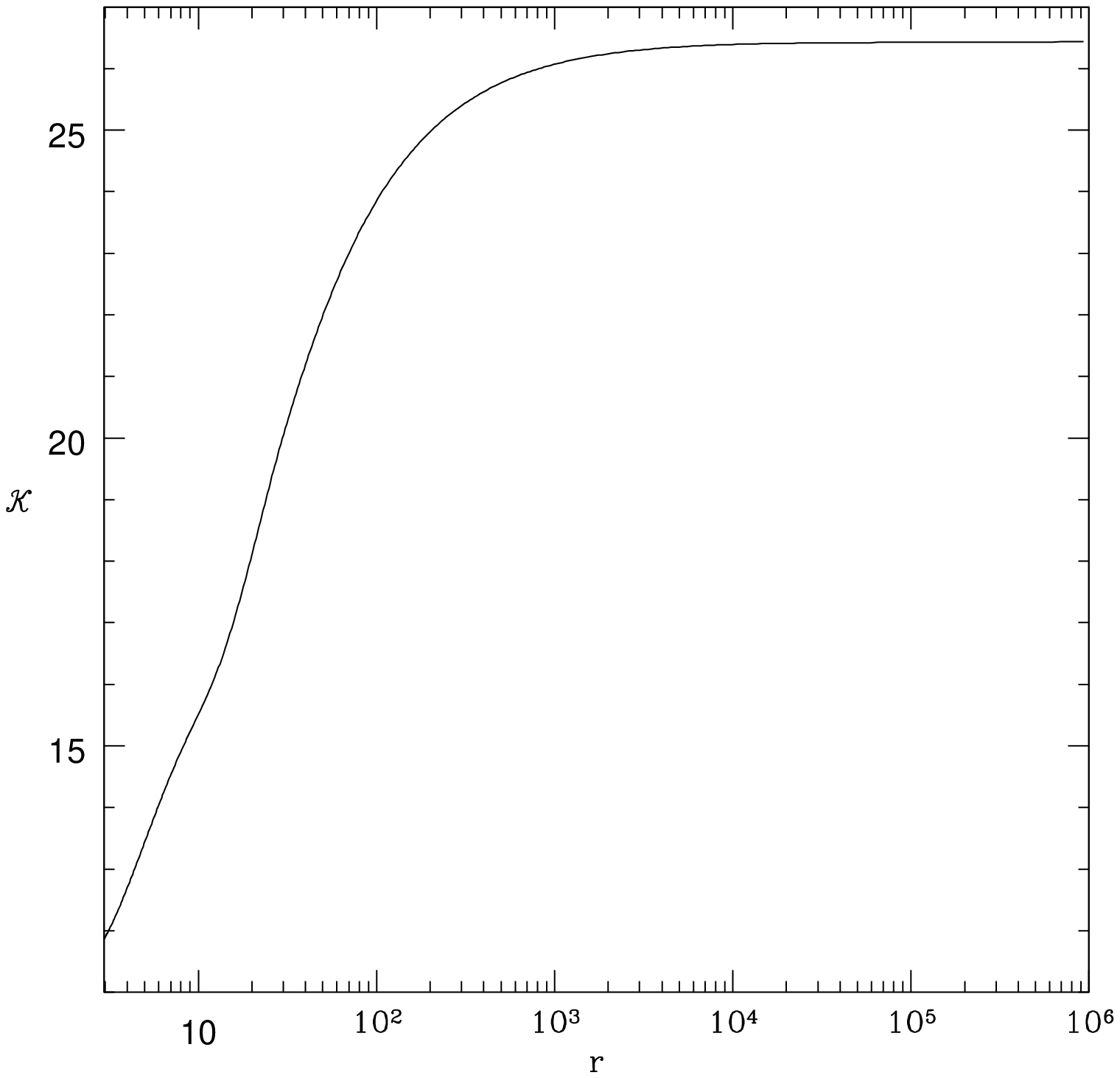}
\vskip -0.1cm
 \caption{Variation of $\cal K$ with $r$ for $\ell=1,~E=1.35$}
\label{lab:figC2}
 \end{center}
\end{figure}

If ${\cal K} \lsim 1$, then the assumption of conical flow geometry will not hold. 
To analyse, we choose $\alpha=10^\circ$ and take an example of A-type solution (Fig. \ref{lab:A_type}), for $\ell=1$ and $E=1.35$, and plot $\cal K$. The variation of $\cal K$ is shown in Fig. (\ref{lab:figC2}). 

We see that $\cal K$ remains significantly greater than $1$ throughout the jet extent and it increases with $r$. Hence we can safely state that the assumption of conical flow geometry is a reasonable one.

\section*{Acknowledgments}

\end{document}